# Exploring PdCrAs Half-Heusler Alloy for Sustainable Energy Solutions: An Ab-initio Study⋆

Rajinder **Singh**[a], Shyam Lal **Gupta**[b], Sumit **Kumar**[c], Lalit **Abhilashi**[a], Diwaker[d,*] and Ashwani **Kumar**[a,**]

[a]*School of Basic Sciences, Abhilashi University Mandi, Mandi, 175045, H P , INDIA*
[b]*Exploring Physics for Interdisciplinary Science and Technology (EPIST) Lab, HarishChandra Research Institute, Prayagraj, Allahabad, 211019, U P , INDIA*
[c]*Department of Physics, Government College, Una, 174303, H P , INDIA*
[d]*Department of Physics, SCVB Government College, Palampur, Kangra, 176061, H P , INDIA*

## ABSTRACT

This work presents a comprehensive investigation of the HH alloy PdCrAs using first-principles methods, highlighting its potential applications in various fields, including spintronics, thermoelectrics, and optoelectronics. We employed density functional theory (DFT) within the full-potential linearized augmented plane wave (FLAPW) framework. Structural optimizations indicate that the alloy stabilizes in the ferromagnetic phase. Both mechanical and dynamical stability have been confirmed through analyses of elastic constants and phonon dispersion. Our calculations of the electronic band structure and density of states (DOS) reveal that PdCrAs exhibits half-metallic behavior, with a spin-polarized band gap of 0.670 eV in the minority spin channel. The magnetic moment aligns with the Slater-Pauling (SP) rule, indicating robust ferromagnetism. Mechanical analysis shows that the material is ductile in nature. Thermodynamic analysis highlights the alloy's resilience, supported by consistent trends in entropy, heat capacity, and Debye temperature. Its optical response demonstrates strong absorption in the visible and ultraviolet (UV) regions, along with pronounced dielectric and plasmonic features, suggesting potential applications in optoelectronics and reflective coatings. Furthermore, evaluations of the transport properties reveal high Seebeck coefficients and a significantly tunable figure of merit (ZT), with values approaching 0.9 across the temperature range of 300-1500 K, indicating excellent thermoelectric characteristics. Overall, these findings position PdCrAs as a promising multifunctional material suitable sustainable energy solutions

## 1. Introduction

Researchers have been doing qualitative research in the clean energy sector over the past 20 years in an effort to find reliable, affordable materials. Rapid industrialization has resulted in a high demand for energy, which eventually drives the consumption of natural resources. As a result, energy consumption needs to be decreased and alternative resources of energy need to be invented. In a recent study it was found that a large amount of energy is lost in the form of heat so we need materials which can perform better in response to light, pressure and temperature. Such materials which have better performance w.r.t temperature are known as thermo electric materials denoted by (TE) [1, 2, 3, 4, 5, 6, 7, 8, 9, 10, 11, 12, 13]. In recent studies it was found that with ZT values of 0.71 and 0.72 at 1200 K, respectively, half-Heusler compounds like FeTaP and FeTaAs have been found to have very advantageous thermoelectric materials [14]. Lead telluride alloy PbTe doped with thallium is another useful thermoelectric material which achieves a ZT of 1.5 at 773K [15]. The highest value of ZT reported is 2.2 when lead telluride is used to convert waste heat into electricity in 20212. Heusler alloys have been found to posses tremendous potential in applications for power generation at higher temperatures. They are known to possess narrow bandgap, high seebeck coefficient, and optimal value of ZT which makes them suitbale in thermoelectric devices. Heusler alloys have a deep structural stratification that includes complete Heusler, half Heusler, and inverse Heusler configurations. Additionally, this categorization broadens its focus to include quaternary and binary compounds, offering a framework for understanding the special physical characteristics of these substances. The chemical formula XYZ denotes a typical half-Heusler compound, where Z stands for a main group element and X and Y for electropositive transition metals or rare-earth elements [16, 17, 18, 19, 20, 21]. The figure of merit (ZT) is used to evaluate a thermoelectric material's effectiveness, which is influenced by multiple factors. The following formula is used to determine the figure of merit:

$$ZT = \frac{S^2 \sigma T}{\kappa} \qquad (1)$$

Here, S stands for the Seebeck coefficient, $\sigma$ for the electrical conductivity, and $\kappa$ for the thermal conductivity of the chosen thermoelectric material in the equation above. A thermoelectric material's figure of merit indicates its performance; larger the figure of merit, better will be material's performance. High electrical conductivity, low thermal conductivity, and a high Seebeck coefficient are desirable properties for thermoelectric materials. Among the well-known thermoelectric materials on the market are silicon-germanium alloys, lead telluride, and bismuth telluride. Skutterudites, half-Heusler (HH) alloys, clathrates, oxy-selenides, chalcogenides, organic hybrid materials, and nanostructures are among several other thermoelectric materials that are presently the subject of investigation.We now focus on half-Heusler compounds, whose multifunctional qualities have been the subject of much research. Excellent thermoelectric materials with high thermoelectric efficiency (ZT) throughout a wide range of operating temperatures are half-Heusler (HH) alloys [22, 23, 24, 25]. Numerous methods, such as doping, substitution, vacancies, defect engineering, and dimension reduction, have been studied to enhance the thermoelectric performance of HH materials. HHs typically consist of one p-block element and two d-block elements. The compound that was produced when nonmagnetic elements were alloyed in HH alloys was discovered to possess magnetic properties. HH alloys have been seen to have half-metallic (HM) properties. This suggests that the substance functions similarly like a metal in one spin orientation and as a semiconductor in another. These unique characteristics result from the specific configuration of atoms inside the materials, their interactions with each other, and the overall number of valence electrons. The high magnetoresistance (HM) property of these alloys is highly appreciated for spintronic devices as spin injectors, spin valves, and magnetic random access memory (MRAM) devices. In this work, we identify Palladium-based Heusler alloys, or PdCrAs, to gain a better understanding of their thermoelectric and spintronic properties..

## 2. Computational Details

DFT is an efficient method for solving the many-body electron problem by applying the Kohn–Sham (KS) equation, which reduces the problem to a single-body electron problem. We employed the full potential linearized augmented plane wave (FP-LAPW) techniques of the all-electron code package WIEN2k to solve the self-consistent single-electron issue. This allowed us to calculate the physical properties of PdCrAs HH from basic principles. To get the ideal lattice parameters, we fitted the energy versus volume data using the Birch–Murnaghan equation. We employed the exchange-correlation (XC) functionals and flexible pseudopotentials from WIEN2K for our self-consistent ground state energy calculations [26, 27]. The most popular XC functional is the generalized gradient approximation (GGA). This paper presents data from the TB-mBJ method which demonstrate improved accuracy in the electrical density of states (DOS) and band structure [28]. Phonon dispersion was calculated using the Phonopy technique with the finite displacement approach. The IRelast package included with WIEN2k was used to assess the mechanical stability of the HH alloy

PdCrAs HH alloy. A constant energy value of -6.0 Ry was utilized to differentiate between valence and core electrons. Following the Monkhorst–Pack approach, a 10x10x10 K-point mesh was used for Brillouin zone sampling. The muffin tin (RMT) radii for each of the relevant atoms were RMT [Pd] = 2.45, RMT [Cr] = 2.39, and RMT [As] = 2.33. The cut-off value RMT × KMAX was set at 7 to limit the quantity of generated plane waves. The GIBBS2 software was used to examine the thermodynamic properties of our materials within the melting point temperature range using the quasi-harmonic approximation. The transport properties were calculated using the BoltzTrap2 code, which operates under the rigid band approximation and assumes a constant relaxation time. We calculated the electrical conductivity, electronic thermal conductivity, and Seebeck coefficient as we increased the temperature from 600 to 1500 K.

## 3. Results and Discussion

### 3.1. Structural properties

Fig. 1. [a] shows the optimized crystal structure and [b] energy vs volume curve for HH alloy PdCrAs as per wyckoff positions given in table 1. In order to examine the structural characteristics of the PdCrAs HH alloy ground state configuration, structural/geometry optimization was initially done by reducing the alloys' total energy in relation no 2. known as to the Birch-Murungnan equation. By examining the volume-energy curve and fitting it with the Birch–Murnaghan equation of state, which is shown below, the ideal lattice parameters for this alloy were identified and are tabulated in table 2.

$$E_{tot}(V) = E_0(V) + \frac{B_0 V}{B_0(B_0 - 1)} \left[ B \left( 1 - \frac{V_0}{V} \right) + \left( \frac{V_0}{V} \right)^{B_0'} - 1 \right]$$

(2)

Since the (FM) states of the PdCrAs HH alloy have the lowest ground state energy, it may be concluded that it is ferromagnetic in nature. [29, 30]. The additional parameters in this work have been calculated using the ferromagnetic states of PdCrAs HH alloy. The stability of HH alloy PdCrAs is confirmed by calculation its formation energy given as

$$\Delta H_f = \frac{E_{tot}(PdCrAs) - (E_{Pd} + E_{Cr} + E_{As})}{3}$$

(3)

where $\Delta H_f$ $E_{tot}$, $E_{Pd}$, $E_{Cr}$ and $E_{As}$ are the energy of formation, the optimal total energy of the PdCrAs compound, the energy of the Pd atom's ground state, the lowest energy state of the Cr atom, and the lowest energy state of the As atom. The investigated compound's structural stability is demonstrated by the negative value of its formation energy, which is approximately -4.92 eV/atom. The formation energy $E_{form}$ from the OQMD database was determined for PdCrAs HH alloy and is found to be -0.259 eV/Atom. This -ve value of formation energy ascertain the possibility of experimental synthesis of PdCrAs HH alloy. Also, the stability values

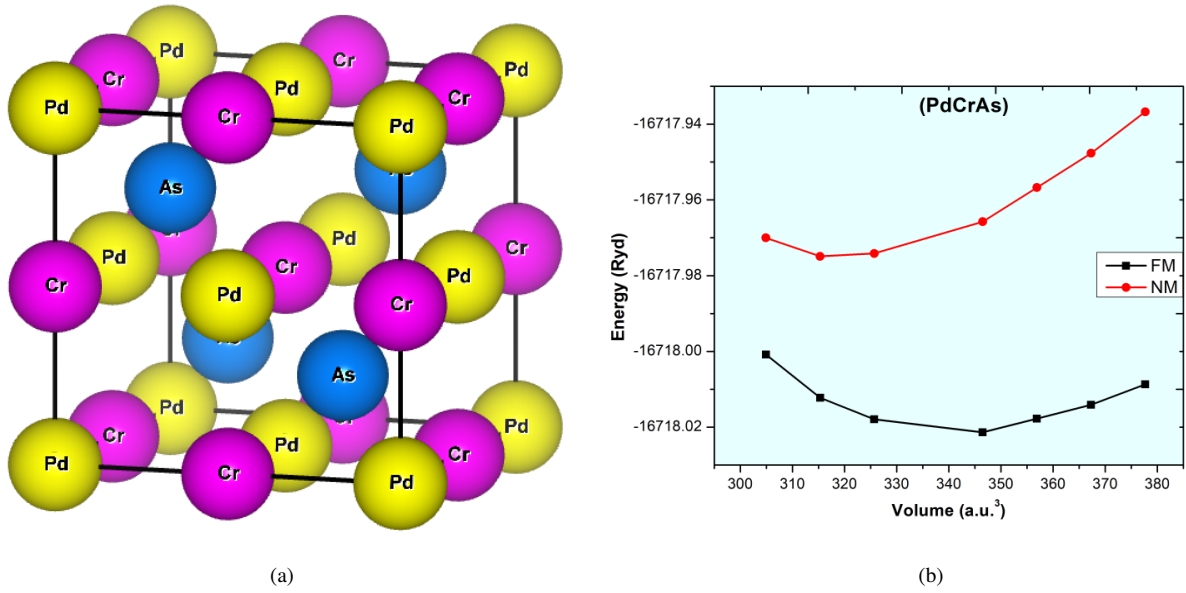

**(a)**                                     **(b)**

**Figure 1:** [a] The crystal structures of the cubic PdCrAs half-Heusler alloy [b] Energy Vs Volume curve for PdCrAs HH alloy

**Table 1**
Wyckoff position of atoms in the unit cell of the cubic half- Heusler alloy PdCrAs.

| Type | Pd | Cr | As |
|---|---|---|---|
| PdCrAs | (0, 0, 0) | (0.5, 0.5, 0.5) | (0.25, 0.25, 0.25) |

are 0.044 eV per atom for the PdCrAs HH Alloy [31, 32]. The HH alloy PdCrAs computed phonon band dispersion and atomic projected phonon density of states are displayed in fig. 2. As shown in fig. 2, the dispersion curves were plotted along the Brillouin zone's high symmetry path. The Phonopy packages were used to do these computations. The spectra show no negative (imaginary) frequencies, supporting the material's dynamic stability in the ferromagnetic phase as tabulated in table 1.

### 3.2. Electronic and Magnetic properties

The band structures of the PdCrAs HH alloy ferromagnetic state are shown in fig. 2, where their band structures were analyzed using the mBJ-GGA approximation. The half-metallic characteristics are present in the band structure with a energy band gap of 0.670 eV near the Fermi level. Within the Brillouin zone, the band structure profile of the PdCAs HH alloy shows a indirect band gap along the Γ direction. The mBJ-GGA approximation was used to compute the partial density of states (PDOS) and total density of states (TDOS) for PdCrAs HH alloy. Fig. 4. illustrates how the band gap in the spin-down arrangement closely matches the one determined by band structure calculations. The spin polarization P given by Eqn. 4. was calculated for the majority and minority bands in order to get quantitative data that would show the half metallic property from the electronic band structure calculations.

$$P = \frac{N_\uparrow(E_F) - N_\downarrow(E_F)}{N_\uparrow(E_F) + N_\downarrow(E_F)} \times 100\% \qquad (4)$$

The findings verify that the HH alloy PdCrAs has 100% spin polarization and attractive properties that make it a viable option for magnetic storage and spintronic systems applications as a sustainable energy source. The purpose of the TDOS and PDOS plot is to give insight into the nature of bonding between the orbitals of the individual atoms. The Cr atom's $t_{2g}$ states predominate in the region below the Fermi energy, whereas the $e_g$ states do so in the region above it, according to the minority-spin channels of the PdCrAs HH alloy. Both states exhibit metallic characteristics because they are occupied around the Fermi energy. The crystal field splitting of the Cr atom's d-orbital into $t_{2g}$ and $e_g$ is the cause

**Table 2**
Optimized structural and other parameters of PdCrAs HH alloy.

| HH Alloy | phase | a(Å) | $V_o$(GPa) | B(Gpa) | B′(GPa) | $E_o$(Ryd) |
|---|---|---|---|---|---|---|
| PdCrAs | NM | 5.7505 | 320.8217 | 155.0919 | 4.8011 | -16717.974826 |
|  | FM | 5.8675 | 340.7985 | 123.7260 | 5.7086 | -16718.021391 |

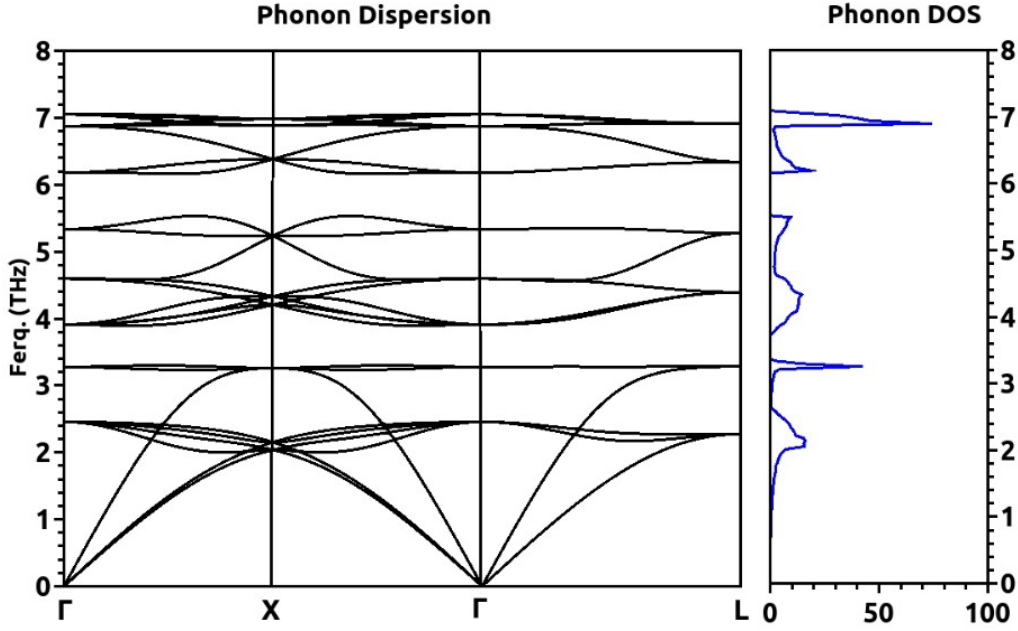

**Phonon Dispersion**

**Phonon DOS**

(a)

**Figure 2:** Phonon dispersion curve of the PdCrAs HH alloy at the equilibrium lattice constant.

of the gaps observed around the Fermi energy in the majority of spin straits for the PdCrAs HH alloy. The depletion of electrons in the majority spin states results in empty states because of the exchange interactions between the Cr d-orbital and the Pd d-orbital. As a result, a half-metallic gap forms around the Fermi energy. Additionally, we saw that the Cr d-orbital and the Pd d-orbitall had bonding states that were slightly below the Fermi energy and anti-bonding states that were slightly above the Fermi energy. The orientation of electron spins and a material's spin magnetic moment (MM) influence its magnetic characteristics. Table 3. tabulates the PdCrAs HH alloy total and partial magnetic moments. In the ferromagnetic state, the calculated magnetic moment of HH alloy PdCRas via mBJ-GGA is an integer ($3\mu B$). Notably, the Cr metal provides the majority of the contribution, while other atoms Pd and As have little contributions. Antiferromagnetic interactions between valence band electrons are indicated by the negative partial magnetic moments. The Slater-Pauling rule (SPR) can be used to determine the total magnetic moment (MT) of a half-metallic (HH) material as indicated by the following equation [33]:

$$M_T = Z_T - N \tag{5}$$

where N can have values of 18, 24, and 28 depending on the electronic structure of the material and the number of atoms in each formula, and $Z_T$ is the total number of valence electrons. For our HH alloy PdCrAs $Z_T$ is 21 (Pd = 10, Cr = 6 and As = 5). In conclusion, the partial magnetic moments estimated for the atomic contributions agree with the magnetic moment value for PdCrAs HH alloy, which is ($3\mu B$). Furthermore, the practicality and usability of our material depend heavily on our understanding of its Curie temperature. The critical temperature and the total magnetic moment of a half-Heusler material have a linear relationship, as suggested by Wurmehl et al [34] and Candan et al [35]. This is as follows:

$$T_C = 23 + 181 M_T \tag{6}$$

The HH alloy PdCrAs estimated Curie temperature is 566 K which is higher than room temperature, which makes it suitable for spintronic applications.

### 3.3. Mechanical stability

We used the IRelast Package intergrated in Wien2k code to compute the elastic constants in order to find out more about phase transitions and structural stability as well as to forecast whether PdCrAs will be stable under stress or not. The stress–strain relationship was used to obtain the elastic constant values $C_{11}$, $C_{12}$, and $C_{44}$ [36, 37, 38] which can be expressed as following

$$C_{11} - C_{12} > 0 \tag{7}$$

$$C_{11} > 0 \tag{8}$$

$$C_{11} + 2C_{12} > 0 \tag{9}$$

$$C_{12} < B < C_{11} \tag{10}$$

$$C_{44} > 0 \tag{11}$$

$$B = \frac{(C_{11} + 2C_{12})}{3} \tag{12}$$

where B, the bulk modulus, defines the stiffness of the material. According to values tabulated in Table 4. , the PdCrAs HH Alloy satisfies the requirements for mechanical stability by displaying positive values for elastic constants [39]. We used the Voigt–Reuss–Hill approximation to determine the mechanical properties in these calculations, and the output values are shown in Table 4 [40].

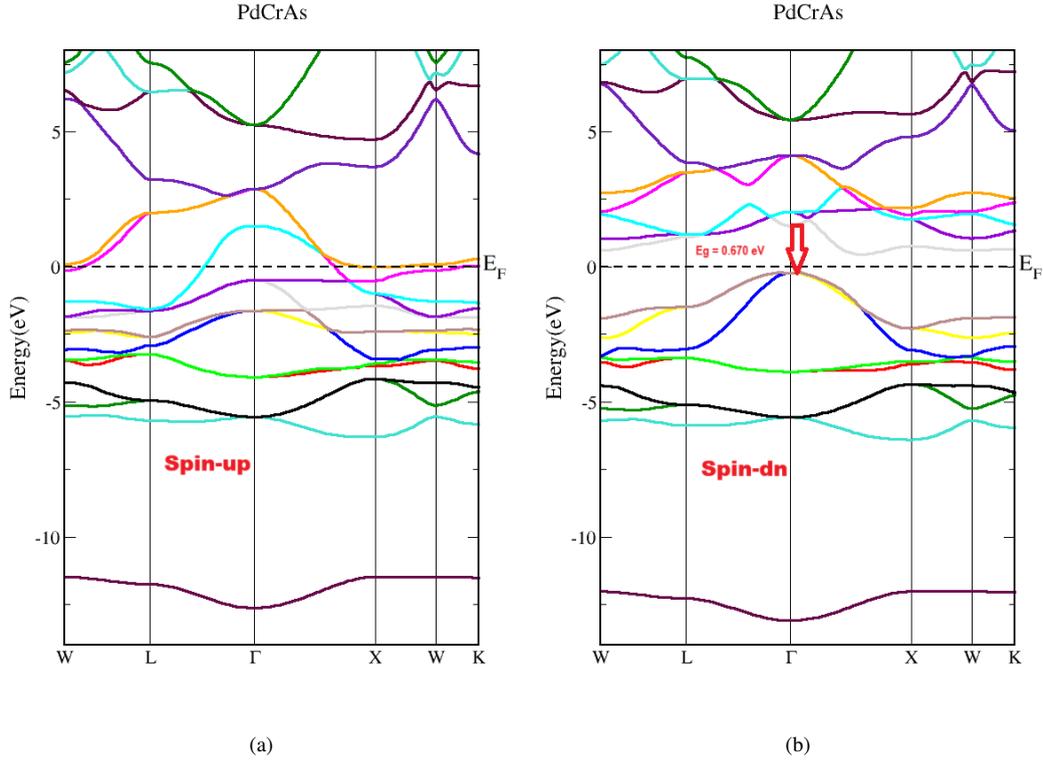

**Figure 3:** Spin-up and spin-down electronic band structure plot of PdCrAs HH alloy using mBJ approach

**Table 3**
Value of Magnetic moment for PdCrAs per unit cell

| | |
|---|---|
| Total $M_T$ ($\mu_B$) | 3.06915 $\mu_B$ |
| MM of Pd ($\mu_B$) | -0.19131 $\mu_B$ |
| MM of Cr ($\mu_B$) | 3.19441 $\mu_B$ |
| MM of As ($\mu_B$) | -0.18962 $\mu_B$ |
| Interstitial MM ($\mu_B$) | 0.25567 $\mu_B$ |

$$Y = \frac{9BG}{3B+G} \quad (13)$$

$$v = \frac{3B-3G}{2(3B+G)} \quad (14)$$

$$G = \frac{G_v + G_R}{2} \quad (15)$$

$$G_v = \frac{(C_{11} - C_{12} + 3C_{44})}{5} \quad (16)$$

$$G_R = \frac{5C_{44} \times (C_{11} - C_{12})}{4C_{44} + 3(C_{11} - C_{12})} \quad (17)$$

$$\zeta = \frac{c_{11} + 8c_{12}}{7c_{11} + 2c_{12}} \quad (18)$$

$$T_m = 553 + 5.911C_{11} \quad (19)$$

where Poisson's ratio, which aids in comprehending the properties of the bonding forces, is represented by the variables $v$, G, $T_m$, $\zeta$, $G_v$, and $G_R$ Young's modulus, which measures the stiffness of the material, Shear modulus, which aids in comprehending the plastic deformation properties of the material, melting temperature, Voigt's and Reuss's shear moduli, respectively, and the Keinman parameter to determine how resilient the material is to stresses brought on by bending or stretching forces [41, 42, 43].

$$A = \frac{2C_{44}}{C_{11} - C_{12}} \quad (20)$$

$$v_M = \left(\frac{1}{3}\right)\left[\frac{2}{u_s^2} + \frac{1}{u_l^2}\right] \quad (21)$$

$$v_s = \sqrt{\frac{G}{\rho}} \quad (22)$$

$$v_l = \sqrt{\frac{(3B+4G)}{3\rho}} \quad (23)$$

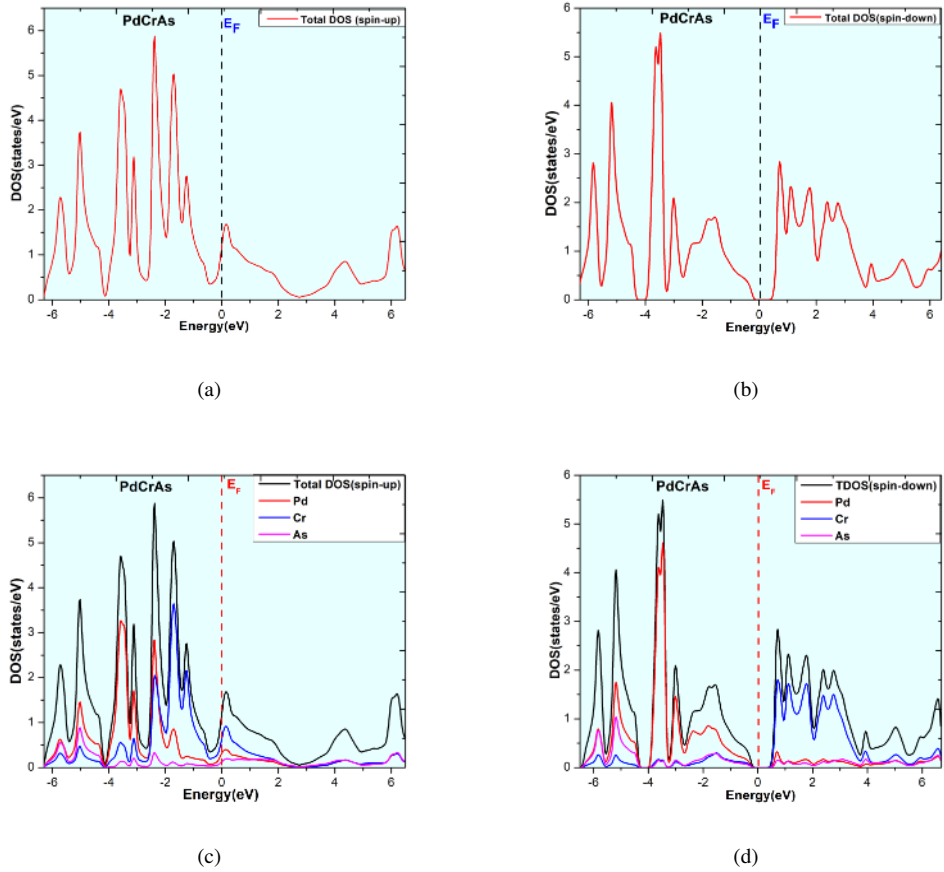

(a)  (b)

(c)  (d)

**Figure 4:** TDOS and PDOS calculated for HH alloy PdCrAs for spin-dn and spin-up configurations TB-mBJ exchange–correlation functionals.

$$H_v = \frac{(1 - 2\nu)E}{6(1 + \nu)} \tag{24}$$

$$\lambda = \frac{E_\nu}{(1 - \nu)(1 + 2\nu)} \tag{25}$$

$$\mu = \frac{E}{2(1 + \nu)} \tag{26}$$

$$\theta_D = \frac{h}{K_B} 3\sqrt{\frac{n^3 N_A \rho}{4\pi M}} V_m \tag{27}$$

According to Pugh, 1.75 is a crucial ratio for distinguishing between brittle and ductile materials [44]. Brittle behavior is implied by a lower B/G ratio, whereas ductile behavior is indicated by a larger ratio. The ductility of the PdCrAs HH alloy combination is indicated by the Pugh ratio (B/G) value in Table 4, which is 2.203 and is greater than 1.75 indicating ductile nature of PdCrAs HH alloy. This is further supported by Cauchy pressure equation given as

$$CP = C_{11} - C_{44} \tag{28}$$

Materials are ductile if the value of CP is positive; otherwise, they are classified as brittle [45]. Our computed value of CP is positive i.e. 118.729 which supports that PdCrAs

HH alloy is ductile in nature. Further information about the bonding forces in the PdCrAs HH alloy can be found using the Poisson's ratio ($\nu$). The presence of a central force is indicated by materials with ($\nu$) values between 0.25 and 0.5. Table 4 shows that the force applied to the material is distributed uniformly, as indicated by the Poisson's ratio of 0.297. The Poisson's ratio ($\nu$) was also used to assess the materials' brittleness and ductility. According to the Frantsevich formula, the crucial Poisson's ratio value for brittle/ductile materials is 0.26. For ductile materials, it is larger than 0.26; otherwise, the material is brittle. The PdCrAs HH alloy's ductile properties are supported by the Poisson's ratio value, which is determined to be more than 0.26 [46]. The Kleinman parameter value of 0.905 sheds light on the PdCrAs material's mechanical properties [47]. A greater Kleinman parameter value (nearing 1) indicates that bond bending reduces the material's resistance to external stress. On the other hand, a low number (around 0) suggests that bond bending barely affects anything. The mechanical strength of the PdCrAs HH alloy is mostly determined by bond bending as opposed to stretching or contracting. Additionally, fig. 5 shows the three-dimensional directional variations for Linear compressibility, Young's modulus (Y), shear modulus (G), and Poisson's ratio ($\nu$). The ELATE code [48] is used in analyzing the directional representation of mechanical properties depending on spatial orientation. A perfect isotropic crystal's three-dimensional surface is projected to be spherical. Any deviation from this shape indicates the existence of some anisotropy. The PdCrAs HH alloy material is therefore categorized as an anisotropic material. Furthermore, the formula given by eqn. no. 19 is used to calculate shear anisotropy which comes out to be 0.362 for our material [48]. It is important to recall that values lower or higher than one indicate anisotropy. Table 4. also displays the different velocities, which were computed by eqn no.(20-22). The microhardness one of the essential property that affects a functional material's applications is given by $H_v$ [49]. Crystals' microhardness can be ascertained by looking at their elastic constants. The Vickers hardness $H_v$ value is shown in Table 5 and is commonly calculated using the formula given by eqn. no. 23. A crystalline solid's Debye temperature ($\theta_D$) [50] is a characteristic temperature that denotes the maximum mode of vibration that atoms can have. It is computed by eqn. no. 26 and value is tabulated in table no. 5. In this equation the Planck's constant, Boltzmann's constant, molecular weight of the material, Avogadro's number, atomic number within the solid's unit cell, and mean sound velocity are represented by the variables h, $k_B$, M, $N_A$, n, and $V_m$, respectively.

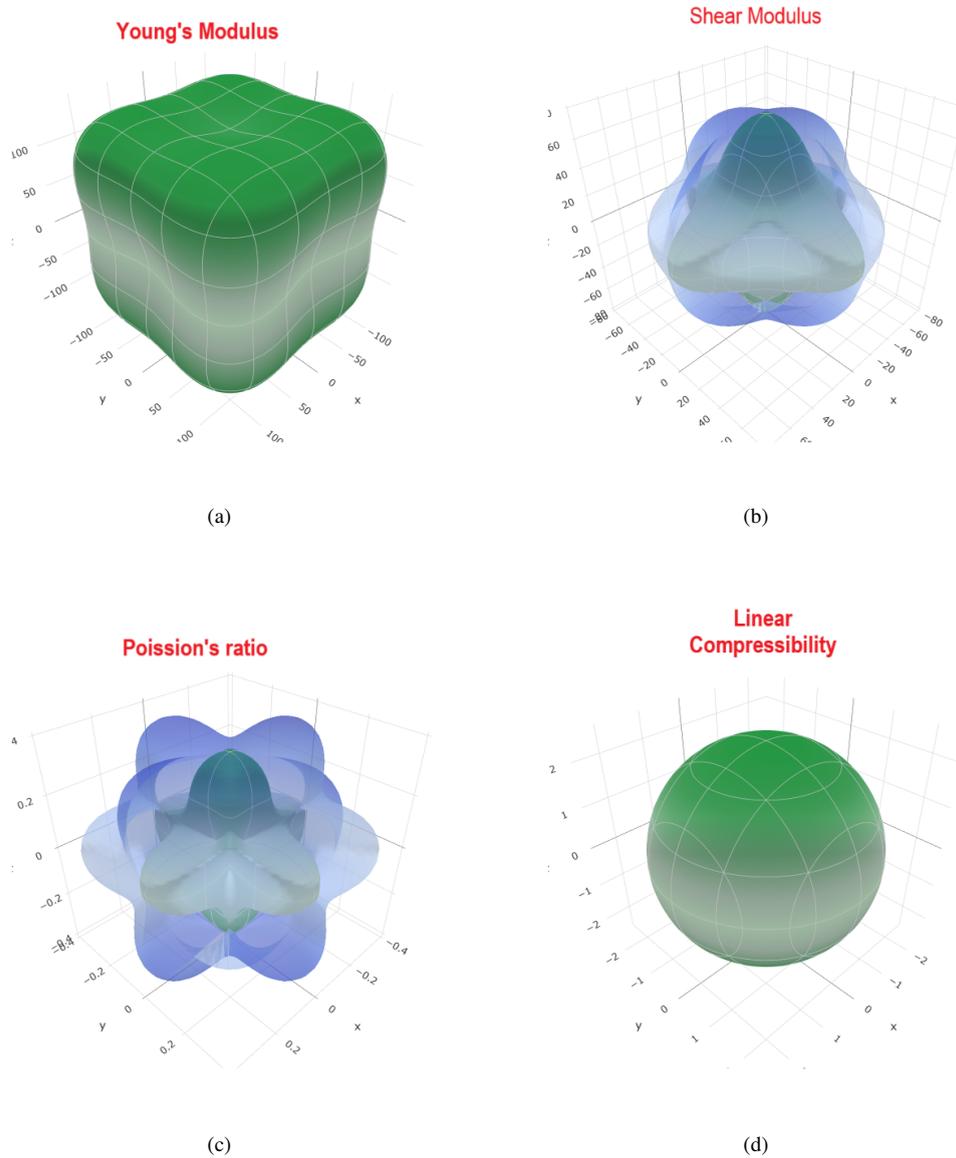

(a)

(b)

(c)

(d)

**Figure 5:** Three-dimensional graphical representation of linear compressibility, Poisson's ratio, Young's, and shear moduli for the mechanically stable PdCrAs HH alloy.

**Table 4**
List of important mechanical parameters of PdCrAS HH alloys

| Parameter | PdCrAs |
|-----------|--------|
| Elastic constant $C_{11}$ (in GPa) | 195.466 |
| Elastic constant $C_{12}$ (in GPa) | 106.264 |
| Elastic constant $C_{44}$ (in GPa) | 76.737 |
| Bulk Modulus B (in GPa) | 135.998 |
| Young's Modulus E (in GPa) | 165.702 |
| Shear Modulus G (in GPa) | 63.882 |
| Pugh's ratio ($\frac{B}{G}$) (in GPa) | 2.203 |
| Anisotropy index A (in GPa) | 0.362 |
| Lame's $1^{st}$ constant $\lambda$ (in GPa) | 94.848 |
| Lame's $2^{nd}$ constant $\mu$ (in GPa) | 61.725 |
| Kleinman parameter $\zeta$ (in GPa) | 0.905 |
| Poisson's Coefficient $\nu$ (in GPa) | 0.297 |
| Melting Temperature $T_m$ (in K) | 1711.55 |
| Transverse Velocity (in m/s) | 2836.114 |
| Longitudnal Velocity (in m/s) | 5333.560 |
| Average Velocity (in m/s) | 3169.038 |
| Debye Temperature $\theta_D$ (in K) | 368.162 |
| Vickers hardness (in GPa) | 5.852 |

## 4. Thermodynamic Properties

The stability, phase transitions, and energy storage capacities of materials are all crucially revealed by the study of their thermodynamic properties. Thus, in the current work using the quasiharmonic Debye model, we examined the heat capacity $C_V$, entropy S, vibrational free energy $F_{vib}$, Debye's vibrational energy $U_{vib}$, bulk modulus B, and Debye temperature ($\theta_D$) for the PdCrAs HH alloy in the temperature range of 0 to 1200 K at 0 GPa of pressure, as shown in fig. 5 and 6. The Debye temperature ($\theta_D$) for our material (fig. 5a) was shown to decrease with increasing temperature, suggesting that the material's melting point and hardness are lesser. In essence, this results from increased volume expansion and bond softening as depicted in fig. 5b. It is important to remember that the Debye temperature is a measurement of thermal conductivity. In other words, a material's thermal conductivity increases with its Debye temperature. At low temperatures, the Debye temperature determined using heat capacity variation is 390 K, which is comparable to the value determined using elastic constants i. e. 369K. The entropy S as a function of temperatures for the HH alloy PdCrAs is shown in fig.5c and is calculated from the following equation given as

$$S = 3K_B \left[ \frac{4}{3} D(\frac{\theta_D}{T}) - ln(1 - exp(\frac{-\theta_D}{T})) \right] \qquad (29)$$

It rises with temperature, suggesting that higher temperatures enable more accessible vibrational states. Particularly at low temperatures, entropy increases quickly as the temperature rises, indicating increased intrinsic disorder and unpredictability in the material. This results from changes in the interatomic force constants, which favor vibrational states that are easier to reach. The obtained value of entropy at room temperature is 75 (J mol$^{-1}$ K$^{-1}$). Heat Capacity $C_V$

is given by

$$C_V = T \left( \frac{\partial S}{\partial T} \right)_V \qquad (30)$$

The heat capacity $C_V$ at a constant volume remained noticeably low at lower temperatures up to 400 K, according to our thermodynamic calculations. At higher temperatures ( greater than 400 K), however, it approaches the Dulong–Petit law (see fig. 5d), indicating that all phonon modes are excited by thermal energy at elevated temperatures [51]. Furthermore, the Debye approximation is seen to govern $C_V$ variation at low temperatures, emphasizing the predominance of quantum processes [52]. It is important to note that only long wavelength acoustic modes are thermally activated at low temperatures; short wavelength (high energy) modes are unlikely to be populated and are therefore not taken into consideration. The heat capacity $C_V$ value obtained at room temperature for HH alloy PdCrAs is $\approx$ 65 (J mol$^{-1}$ K$^{-1}$). Fig. 5e shows the fluctuations of the Debye's vibrational energy $U_{vib}$ and the vibrational free energy $F_{vib}$ of PdCrAs HH alloy with respect to temperature. At 280 K, $U_{vib}$ is 23.02 kJ mol$^{-1}$ K$^{-1}$, whereas $F_{vib}$ is found to be around 1.45 kJ mol$^{-1}$ K$^{-1}$. According to the thermodynamic relationship $F_{vib} = U_{vib}$ -TS, the computed value of free energy for the HH alloy PdCrAs is observed to decrease with temperature. However, as the temperature rises, $U_{vib}$ really rises because (see fig. 5e) of the clear connection between the molecules' increased kinetic energy and heat absorption. Furthermore, the system's higher entropy S is reflected in the vibrational contribution to $F_{vib}$ decreasing with temperature. A higher temperature means more phonon modes available, increasing the number of potential micro states for the system and, as a result, its disorder. This has a clear connection to the basic thermodynamic theory that a system's total entropy rises with temperature. As the temperature rises, the bulk modulus B as seen in fig. 5f gradually decreases. It is true that temperature causes vibrations in the crystal, which change the bonding strength and stiffness. A slow decrease in B is then seen as the temperature rises. This confirms the well-known inversion relationship between bulk modulus and volume expansion and shows how sensitive bulk modulus and volume changes are. Nonetheless, it is seen that the bulk modulus stays constant at temperatures $\approx$ 800 K, which is highly consistent with expansion fluctuation within the same temperature range.

## 5. Optical Properties

The mBJ-GGA approximation was used to determine the optical parameters, which include the dielectric function $\epsilon(\omega)$, refractive index n($\omega$), extinction coefficient k($\omega$), optical conductivity $\sigma(\omega)$, light absorption $\alpha(\omega)$, energy loss spectra L($\omega$), and reflectivity R($\omega$), in the energy range of 0–14 eV. Following relations as below has been used to determine these parameters

$$\epsilon(\omega) = \epsilon_1(\omega) + i\epsilon_2(\omega) \qquad (31)$$

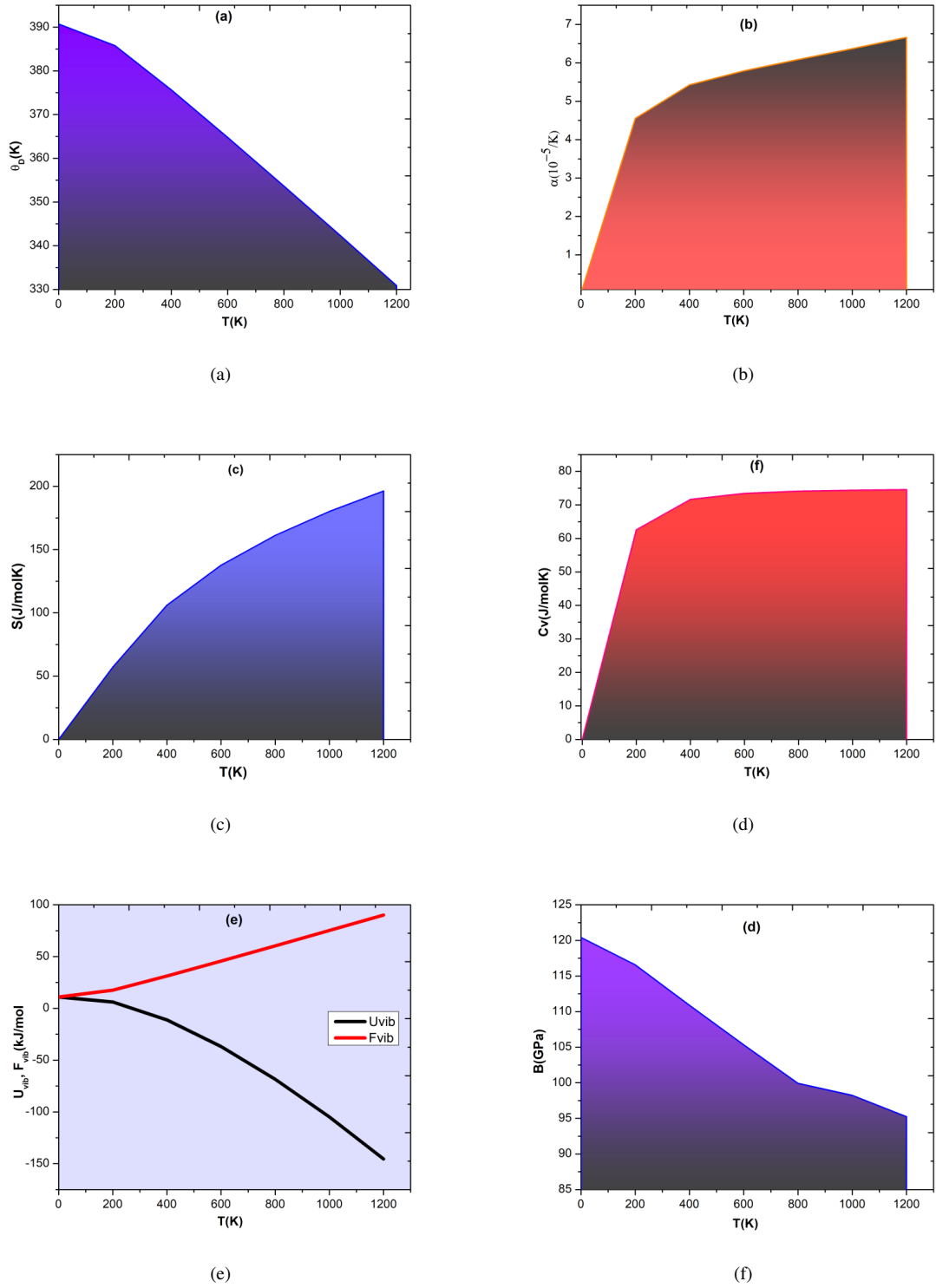

**Figure 6:** Temperature variation of Debye temperature $\theta_D$, expansion coefficient $\alpha$, entropy S, heat capacity $C_V$, vibrational free energy $F_{vib}$ Debye vibrational energy $U_{vib}$, and Bulk modulus B for the PdCrAs HH alloy

$$n(\omega) = \sqrt{\frac{\epsilon_1(\omega)}{2} + \frac{\sqrt{\epsilon_1^2(\omega) + \epsilon_2^2(\omega)}}{2}} \qquad (32)$$

$$K(\omega) = \sqrt{\frac{-\epsilon_1(\omega)}{2} + \frac{\sqrt{\epsilon_1^2(\omega) + \epsilon_2^2(\omega)}}{2}} \qquad (33)$$

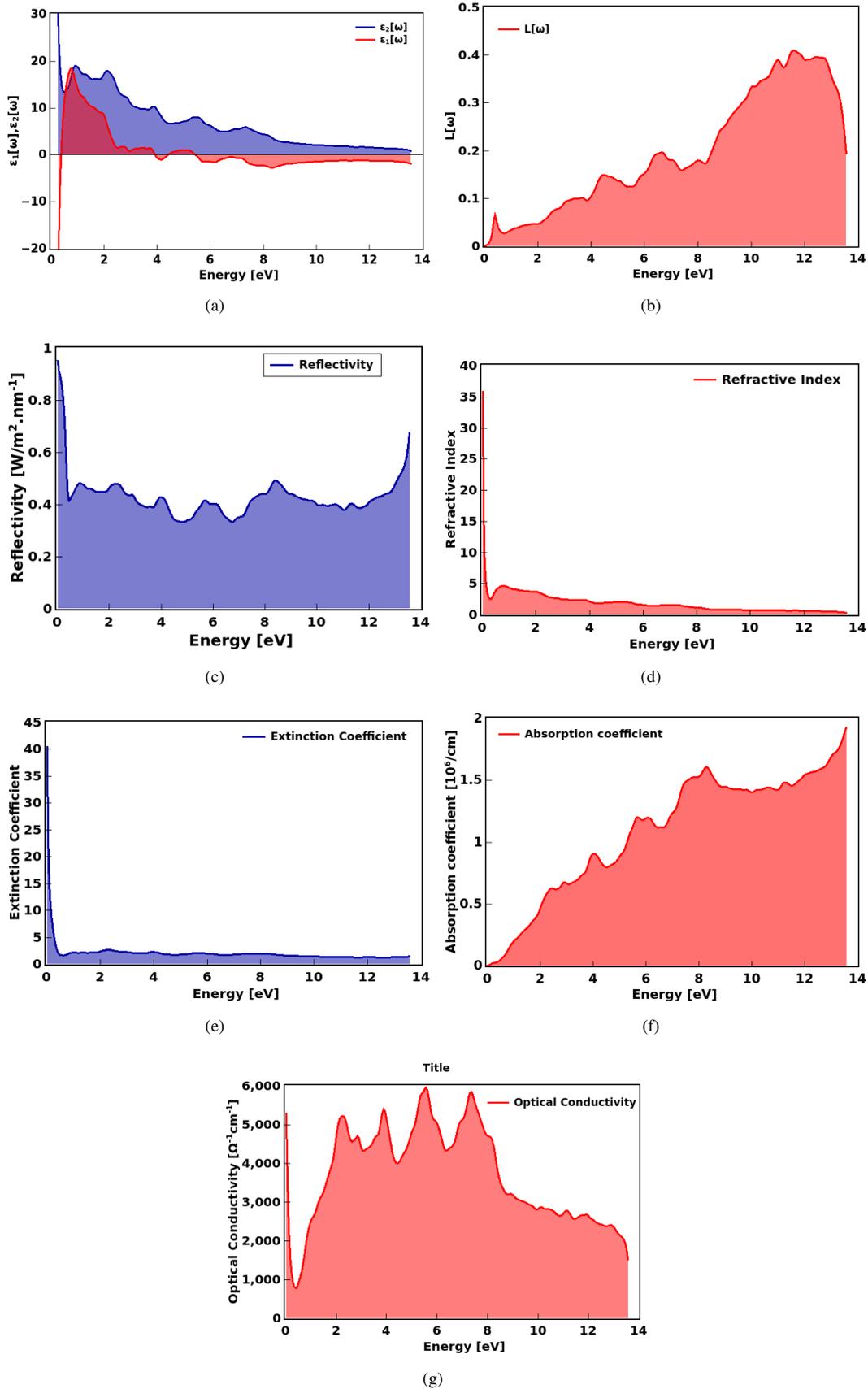

**Figure 7:** (a)–(g) Variation of real and imaginary parts of the dielectric function, energy loss, reflectivity, refractive index, extinction coefficient, light absorption and optical conductivity as a function of photon energy for the HH alloy PdCrAs

$$\sigma(\omega) = \frac{\omega}{4\pi}\epsilon_2(\omega) \quad (34)$$

$$\alpha(\omega) = \frac{4\pi K(\omega)}{\lambda} \quad (35)$$

$$L(\omega) = \frac{\epsilon_2(\omega)}{\epsilon_1^2(\omega) + \epsilon_2^2(\omega)} \quad (36)$$

$$R = \frac{(n-1)^2 + k^2}{(n+1)^2 + k^2} \quad (37)$$

In these equations $\epsilon_1(\omega)$ and $\epsilon_2(\omega)$ are real and imaginary parts of dielectric function which can be computed by the following equations as

$$\epsilon_1(\omega) = 1 + \frac{2}{\pi}P\int_0^\infty \frac{\omega'\epsilon_2(\omega')}{\omega'^2 - \omega^2}d\omega' \quad (38)$$

$$\epsilon_2 = \frac{4\pi e^2}{m^2\omega^2}\sum_{ij}\int \langle i|M|j\rangle I^2 f_i(1 - f_i)\delta(E_f - E_i - \omega)d^3k \quad (39)$$

It's interesting to note how closely the band structure and optical characteristics are related. It's important to understand that critical points play a significant role in peak creation and structural occurrence in the material's optical properties, particularly its dielectric function, optical conductivity, and absorption coefficient. The most notable peaks in the $\epsilon_2$ spectra for the HH alloy PdCrAs, as determined using the mBJ-GGA approximation, are displayed in fig. 7a. At the beginning $\epsilon_1$ starts with negative values and approaches zero at an energy of 0.4eV. The noticeable peak of $\epsilon_1$ is located at, about 1.0eV, as seen in fig. 7a. As photon energy rises, the latter peak falls. A diminished space charge polarization effect is the cause of the notable drop in $\epsilon_1$. In other words, because the input photon frequency is higher than the resonant one, electrons are too slow and scarce to react. For clarity, the dipole's moment opposes the electromagnetic radiation's electric field since the incoming wave and electron motions are out of phase. In areas where $\epsilon_1$ is negative, the material acts like a metal and absorption is commonly seen in these regions. Fig. 7b shows the spectrum of energy loss L($\omega$). it shows a maximum peak value at 11.9eV. Plasmon excitation are frequently connected to them. The presence of collective electron oscillations within the material is the source of these plasmon energy. Furthermore, the resonant peaks in the $\epsilon_2$ spectrum can be correlated with the features seen in both $\epsilon_1$ and R($\omega$) as shown in fig.7c. Resonant peaks are associated with transitions that involve the material's photon absorption. The reflectivity sharply decreases in fig. 7c, which is indicative of the plasma frequency, which is in fact detected at 0.1 eV. Recall that the examination of $\epsilon_2$ also revealed a sharp increase around 0.1 eV, indicating the possibility of an inter band transition [53, 54]. Fig. 10d shows the refractive static index's spectrum fluctuations. It is noted that an intraband conduction type is indicated by a relatively high refractive index at zero energy. Due to negative values of the real dielectric constant, a decrease in the refractive index is seen as the photon energy increases. Observed peaks for the refractive index and $\epsilon_1$ are associated with neighboring photo excitation processes (resonances). The reflectance of the HH alloy PdCrAs varies with incident photon energy, as seen in Fig. 7c. When valence electrons oscillate out of phase with the incoming light, an induced polarization current is produced, which leads to reflectivity. The PdCrAs material exhibits high reflection (91.25%) at zero photon energy (hν = 0). The compound's reflectance decreases to about 45%. This steepness of the discontinuous drop in reflectance identifies the collective oscillations of electrons and is controlled by the relaxation time. Most significantly, the reflectance spectrum shows tiny peaks that are either the result of weak interband transitions or an indication of exciton-induced absorption. Current investigation reveals that Half-Heusler PdCrAs alloy have a wide spectrum range of reflectivity, making them suitable for use as mirrors and reflectors or as coatings. The change in the imaginary part of the refraction index K (extinction coefficient) is displayed in fig. 7e. Due to the material's metallic nature and negative dielectric constant, n and K are both significant at low frequencies. Extinction coefficient sharply drops as the frequency rises until 0.2 eV, at which point it abruptly rises, indicating an interband transition. When the energy is further increased, the extinction coefficient becomes constant over a range of energy upto 11 eV. The efficiency with which a substance absorbs energy from an electromagnetic wave when it strikes it is indicated by the absorption coefficient given by eqn. 35. The absorption coefficient is plotted in fig. 7f. It is noted that the materials exhibit substantial energy absorption across a wide spectral range, spanning from (0-13)eV. Remember that where $\epsilon_1(\omega)$ and K($\omega$) are high (i.e., they represent the electromagnetic dissipation energy into the medium), the predicted absorption is significant. Reflectivity (see Fig. 10c) is negligible in that area because light that is not reflected is instead absorbed by flawless metals. Lastly, the findings indicate that HH alloy PdCrAs may be appropriate for absorption throughout a wide spectral range. Optical conductivity is displayed in fig. 7g. There is static decrease in its value from 3500 $\Omega^{-1}cm^{-1}$ which is characteristic of intraband free electron conduction. For HH Alloy PdCrAs the highest conductivity is observed at 5.9 eV which is 5900 $\Omega^{-1}cm^{-1}$. This peak denotes the material's effective light absorption, which is connected to the electrical transformation that takes place when photons are absorbed. Accordingly, half-Heusler PdCrAs is a promising candidate for optoelectronic applications.

## 6. Transport properties

In this section we have comprehensively studied the thermoelectric response of PdCrAs half Heusler alloys and their dependency using BoltzTraP code. To investigate the different aspects in the domain of thermoelectric (TE) properties, including the figure of merit (ZT), power factor (PF), Seebeck coefficient (S), electrical conductivity and thermal conductivity relative to the constant relaxation time is accomplished as functions of the chemical potential ($\mu - E_F$ [Ha]) at temperatures range 300K, 600K, 900 K, 1200K and 1500K. The chemical potential($\mu - E_F$ [Ha]) is adjusted within the range of $\pm 2.5$ eV. This range is broad enough to reveal the various parameters of transport characteristics relative to chemical potential ($\mu - E_F$ [Ha]) and hence leads to a better understanding of thermoelectric behavior of materials under study. Seebeck coefficient is one of the essential characteristics in thermoelectric devices and materials [55]. It measures a substance's ability to transform a temperature differential into an electric voltage and gives important details on the predominant charge carrier type in the system: p-type conductivity is indicated by a positive value of S, whereas n-type conductivity is indicated by a negative value. The change in S as a function of chemical potential ($\mu - E_F$ [Ha]) at various temperatures is shown in fig. 8a. The appearance of two separate peaks in S, which indicate the presence of both p-type and n-type regimes, is a notable characteristic seen in the given plot. Additionally, the findings unequivocally show that S decreases monotonically as temperature rises, confirming the known negative relationship between temperature and thermoelectric performance. Now, $\frac{\sigma}{\tau}$ is precisely analysed with ($\mu - E_F$ [Ha]) by executing the constant relaxation time method as shown in fig. 8b. We found that $\frac{\sigma}{\tau}$ displays the steady performance across the temperature range. Explicitly, it shows a growth with increasing ($\mu - E_F$ [Ha]) and consequently attributed to the mobility of charge carriers and hence electrical conductivity $\sigma$. At 300 K, critical values of chemical potential ($\mu - E_F$ [Ha]), where $\frac{\sigma}{\tau}$ reaches its maximum value in PdCrAs are -0.1eV and 0.1 eV respectively. Although, attaining these critical values of chemical potential ensures significantly higher doping levels. The electronic component of thermal conductivity normalized by relaxation time $\frac{\kappa_e}{\tau}$ has been systematically assessed at various temperatures as a function of chemical potential relative to the Fermi level ($\mu - E_F$ [Ha]), as shown in fig. 8c. A lower $\frac{\kappa_e}{\tau}$ is preferred for improved thermoelectric efficiency. According to the data, $\frac{\kappa_e}{\tau}$ reaches its lowest value at 300 K and continues to fall as the temperature drops, indicating that 300 K is the ideal temperature for optimizing thermoelectric performance in the compounds under investigation. Notably, the $\frac{\kappa_e}{\tau}$ values are almost insignificant within the range of $\pm 0.1$ eV about ($\mu - E_F$ [Ha]).

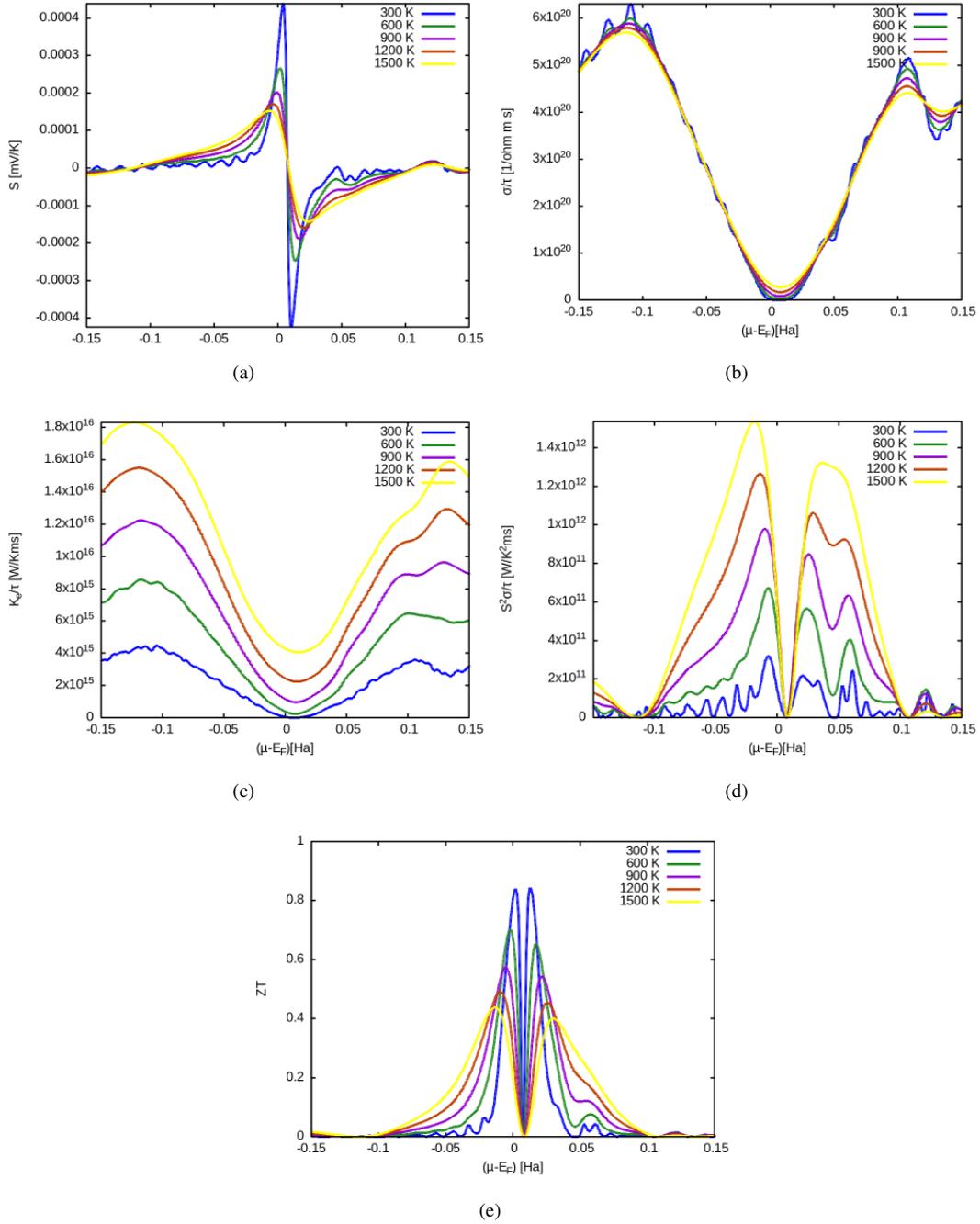

**Figure 8:** Variation of thermoelectric parameters as a function of chemical potential ($\mu - E_F$ [Ha]) at different temperatures [a] Seebeck coefficient [b] Electrical conductivity per relaxation time [c] thermal conductivity per relaxation time [d] Power Factor [e] Figure of Merit ZT

Notably, the $\frac{\kappa_e}{\tau}$ values are almost insignificant within the range of ±0.1 eV about ($\mu - E_F$ [Ha]), indicating the energy window where the alloys are anticipated to function most efficiently. Additionally, the maximal Seebeck coefficient values fall within this region, supporting the synergistic enhancement of thermoelectric performance. Furthermore, a tendency like that of the electrical conductivity $\frac{\kappa_e}{\tau}$ may be seen in the fluctuation of $\frac{\kappa_e}{\tau}$ with respect to ($\mu - E_F$ [Ha]). For all temperatures studied, $\frac{\kappa_e}{\tau}$ shows a slow rise with increasing ($\mu - E_F$ [Ha]), which is in line with the trend shown in $\frac{\sigma}{\tau}$. These results verify that, within the given range of energy and chemical potential, the investigated alloy have exhibit favourable thermoelectric properties, at ambient condition of temperature (300 K). Fig. 8d illustrate that material with greater Seebeck coefficients in the p-type region also have higher power factors (P.F.) in the same domain, and vice versa [56]. At 300 K, the highest values of P.F. values are located at –0.05eV (p-type) and +0.05 eV (n-type) for PdCrAs alloys. The dimensionless figure of merit (ZT) signifies as ZT is one of the fundamental parameters for evaluating the TE efficiency of a material [57] and is given by

$$ZT = \frac{S^2 \sigma T}{\kappa} \qquad (40)$$

Here S is the Seebeck coefficient, $\sigma$ is the electrical conductivity, $\kappa$ refers to total thermal conductivity, which includes both electronic ($\kappa_e$) and lattice ($\kappa_l$) contributions and is given as

$$\kappa = \kappa_e + \kappa_l \qquad (41)$$

T is the absolute temperature. In the numerator $\sigma$ is power factor and this expression captures the interplay between the induced thermoelectric voltage and charge carrier movement. The variation in power factor (PF) arises due to antagonistic dependence of the electrical conductivity ($\sigma$) on Seebeck coefficient (S) and chemical potential. Since the electronic density of states has a significant energy dependency, the Seebeck coefficient is greatest at low carrier density, but electrical conductivity prefers greater carrier concentrations as

$$\sigma = ne\mu \qquad (42)$$

where e is the elementary charge, n is the charge carrier concentration, and $\mu$ is mobility of charge carrier. At high temperatures, the thermal fluctuations can increase n and further elevate ($\sigma$), although it typically occurs at the expense of decreasing S. The detailed thermoelectric behaviour is evident of PdCrAs alloys is plotted in imminent figures. It is observed that at intermediate to high temperatures, the power factor is generally improved because the Seebeck coefficient rises while the electrical conductivity somewhat decreases because of higher carrier scattering. In parallel, phonon–phonon scattering causes the lattice thermal conductivity to drastically decrease with temperature, which lowers $\kappa$. There is constant increase in the figure of merit ZT, which reaches notable levels at high temperatures. The rise in ZT in thermoelectric materials is caused by several processes as shown in fig. 8e. Phase reduction, strain reduction, microstructures (including nanostructures), dopant resonant states, structural order enhancements, heavy hole band exploitation, doping-based sample composition optimization, heavy atomic mass utilization, and high band degeneracy exploitation are some of these developments. These methods include tuning S through low energy carrier filtering and the creation of resonant states close to the Fermi level, enhancing $\sigma$ using metallic phase nanostructures, and reducing $\kappa$ by abruptly hot-pressing bulk materials to reduce grain size to the nano range. When taken together, these tactics work in concern to improve the ZT and advance the potential of thermoelectric materials for a range of uses. In summary, we can achieve (ZT) values approaching 1.0 over a broad temperature range. Its performance at high temperatures shows its potential in energy conversion devices, such as thermoelectric generators and waste heat recovery systems.

## 7. Conclusion

In summary, our comprehensive study of the PdCrAs half-Heusler alloy highlights its potential as a promising candidate for sustainable energy solutions. Our findings indicate that the PdCrAs half-Heusler alloy is stable in the ferromagnetic (FM) phase. Mechanical assessments demonstrate that the alloy is ductile, as supported by calculated mechanical parameters. Phonon dispersion relations show no negative frequencies, and the enthalpy of formation confirms the compound's dynamic stability and suitability for experimental synthesis. The analysis of the electronic structure reveals that PdCrAs exhibits half-metallic behavior, with a significant semiconducting band gap of 0.670 eV in the minority spin channel while maintaining metallic character in the majority spin channel. The density of states (DOS) analysis further supports this finding, indicating the contribution of the d-block element vanadium (V) near the Fermi level, which enhances its half-metallic nature. The calculated total magnetic moment aligns perfectly with the Slater-Pauling rule, confirming the alloy's half-metallic ferromagnetic nature with 100% spin polarization at the Fermi level. The thermodynamic properties of the alloy confirm its thermal stability and heat capacity behavior under varying temperature conditions. The observed decrease in Debye temperature with increasing temperature suggests strong thermal resistance, making it suitable for high-temperature applications. Analysis of heat capacity and entropy trends indicates a stable thermodynamic response. Furthermore, the optical properties reveal strong interband transitions and high absorption in the infrared (IR) and ultraviolet (UV) regions, along with characteristics beneficial for plasmonic and reflective technologies. Importantly, the thermoelectric study indicates high and temperature-dependent ZT values, reaching approximately 0.9 between 300 K and 1500 K. This underscores the potential of the PdCrAs half-Heusler alloy in energy conversion applications. Overall, this work establishes PdCrAs as a promising candidate for integration into next-generation spintronics, optoelectronic devices, and thermoelectric applications.

## Declaration of Competing Interest

The authors declare that they have no known competing financial interests or personal relationships that could have appeared to influence the work reported in this paper.

## Acknowledgement

Authors gratefully acknowledge the support by Department of Higher Education, Government of Himachal Pradesh, Shimla.

## CRediT authorship contribution statement

**Rajinder Kumar:** Data compilation, Writing - original draft Writing - review and editing. **Shyam Lal Gupta:** Conceptualization, Methodology, Data curation, Writing - original draft Writing - review and editing. **Sumit Kumar:** Software, Workstation, Data generation. **Lalit Abhilashi:** Software, Workstation, Data generation. **Diwaker:** Conceptualization, Methodology, Data curation, Writing - original draft Writing - review and editing. **Ashwani Kumar:** Data compilation, Writing - original draft Writing - review and editing.